\documentclass{aipproc}
     
\usepackage{epsfig}
     
\begin{document}
     
\title{ A Feynman Diagram Analyser DIANA -- \\Graphic Facilities }
\author{J.~Fleischer and M.~Tentyukov\thanks{Supported by DFG under FL241/4-1
and in part by 
RFBR $\#$98-02-16923.}\phantom{a}\thanks{On leave from BLTP JINR, Dubna, Russia}  
}    
\affiliation{Fakult\"{a}t f\"{u}r Physik, Universit\"{a}t Bielefeld   
D-33615 Bielefeld}
     
\begin{abstract}
New developments concerning the extension of the recently introduced \cite{DIANA}
Feynman diagram analyser DIANA are presented.
\end{abstract}

\maketitle

       
\begin{figure}[!tb]
\vspace{-0.5cm}
\epsfysize=90mm \epsfbox{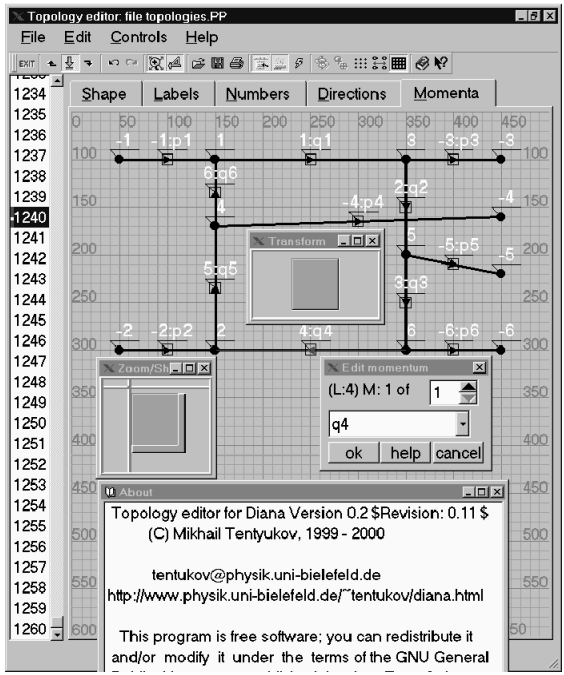}
\vspace{-1.5cm}
\caption{\label{tedi} Topology editor.}
\end{figure}
       
\begin{figure}[!tb]
\vspace{-.5cm}
\epsfysize=110mm \epsfbox{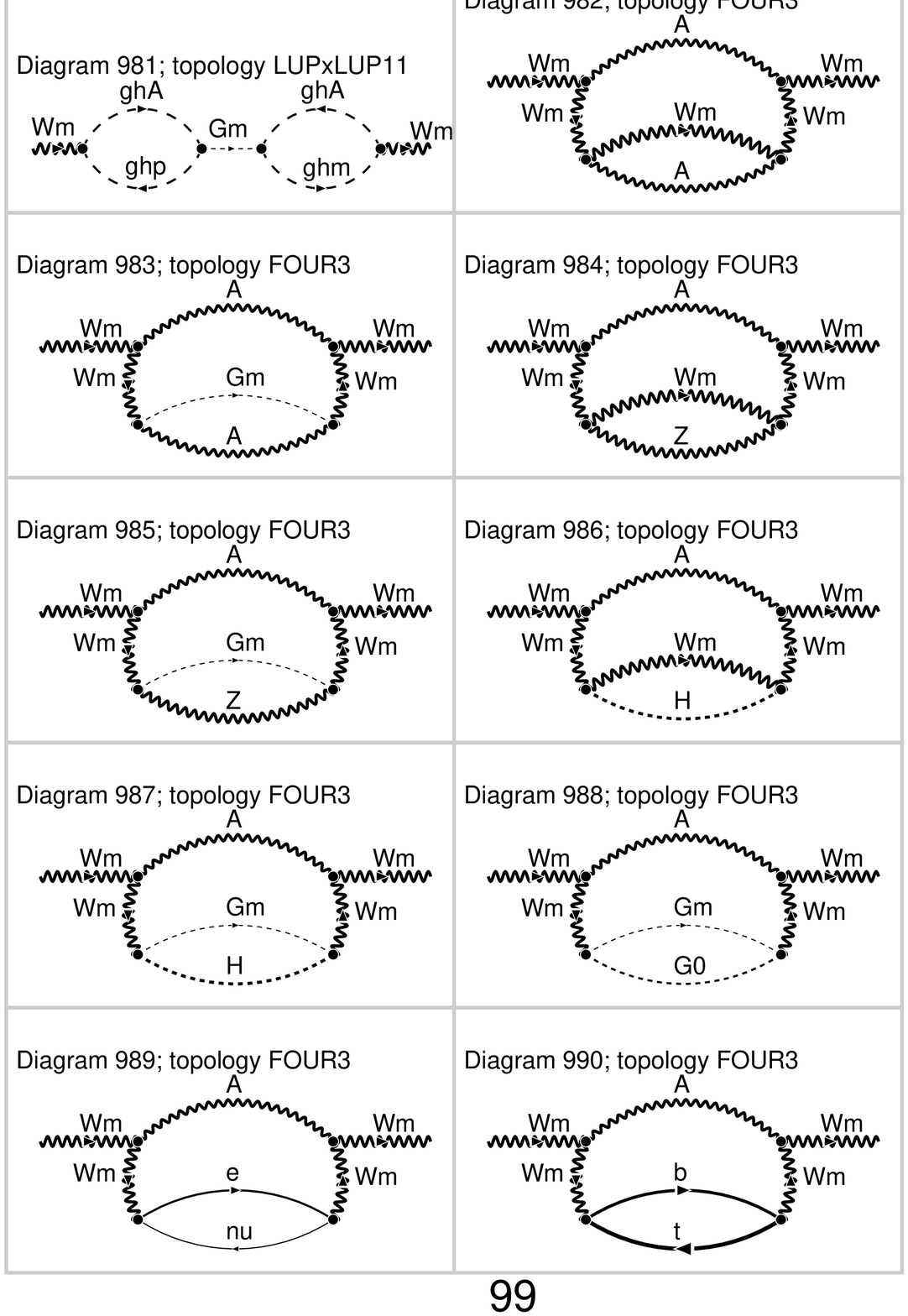}
\vspace{-1.5cm}
\caption{\label{firstPS} Sample of a page with diagrams arranged in columns and rows.}
\end{figure}
                                                               
\begin{figure}[!tb]                                            
\vspace{-0.5cm}                                                
\epsfysize=110mm \epsfbox{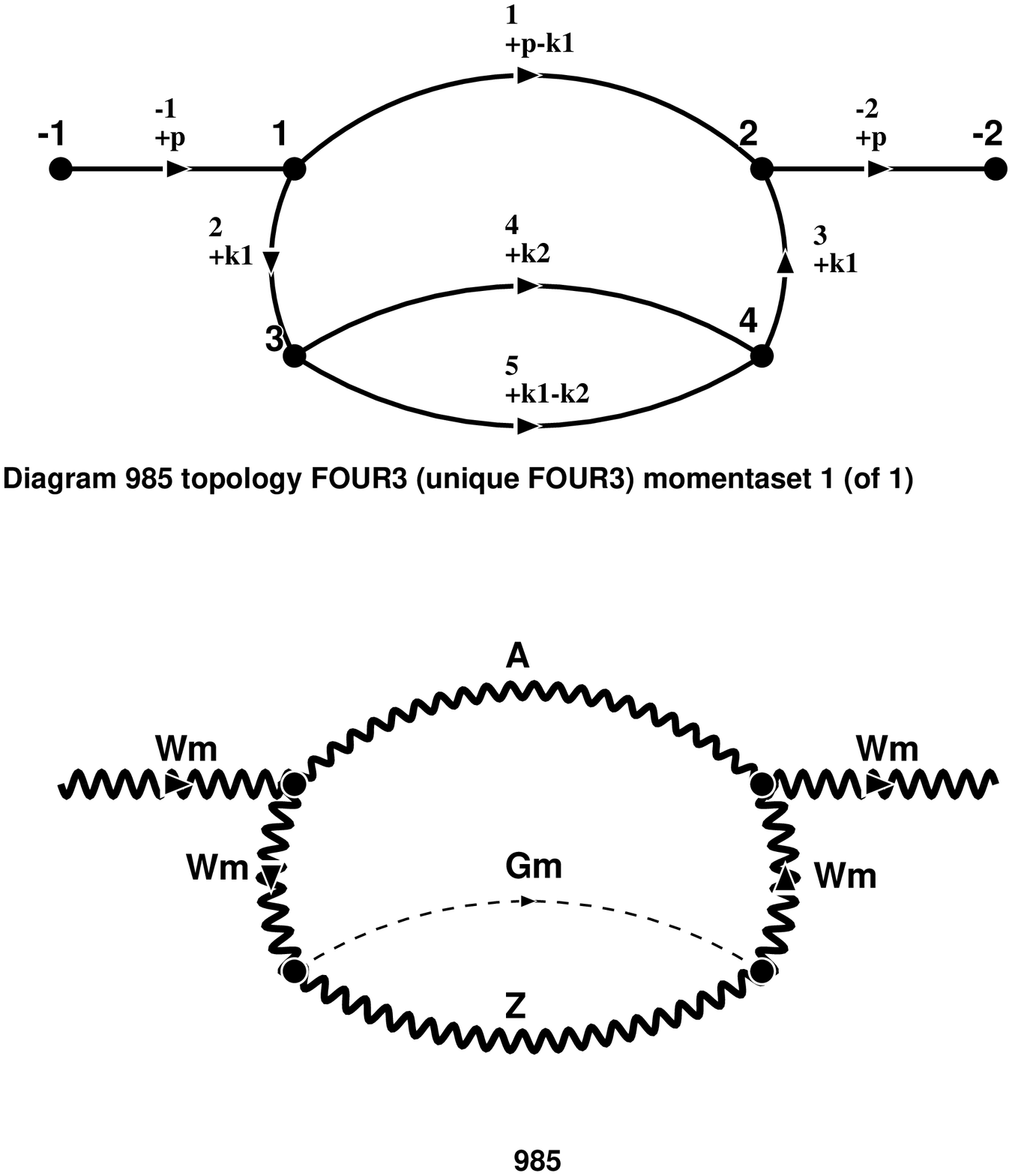}                       
\vspace{-1.5cm}                                                
\caption{\label{secondPS} Details of one diagram.}
\end{figure}      
       
   Recent high precision experiments require, on the side of the
   theory,
high-precision calculations resulting in the evaluation of higher
loop dia\-grams in the Standard Model.
For specific processes thousands of multiloop Feynman
dia\-grams do contribute.
 Of course, the contribution of most of these
diagrams is very small. But sometimes it is not so easy to distinguish
between important and unimportant diagrams. On the other hand, we
often need to take into account all diagrams, to verify gauge
independence, or cancellation of divergences.
It turns out impossible to perform
these calculations by hand. This makes the request for automation a
high-priority task.
       
Our aim is to create a universal
software tool for piloting the process of generating the source
code  in multi-loop order
for analytical or numerical evaluations  and to keep the control of
the process  in general. Based on this instrument, we can attempt to
   build              
a complete package performing the computation of any given process
in the framework of any concrete model.
                      
   The project called DIANA (DIagram ANAlyser) \cite{DIANA}
 for the evalua\-tion of Feynman diagrams was started by our group 
some time  ago.  At present, the core  part is finished. The recent
development of  this project  will be shortly described below.
       
DIANA has been developed for the analytic evaluation of Feynman 
diagrams in terms of computer algebra packages, for which we use FORM
\cite{FORM} , 
but which can in principle be substituted by another language. 
The user has to prepare a file, which contains the model
and process specifications, see details in \cite{DIANA}. 
Reading this file, DIANA will generate all
necessary other files and then invoke the topology editor. 
The purpose of
the topology editor is to make the shapes for the topologies and
to introduce proper integration momenta for the various topologies, 
Fig.  \ref{tedi}.
It is a graphical program written in C++ using the Qt widget library.
For the description of the topology editor see the WEB page\\
\centerline{\footnotesize                                     
\sf                                       
\mbox{http://www.physik.uni-bielefeld.de/\~{}tentukov/topeditor.html}
}      
        
\begin{figure}[!tb]
\vspace{9pt}     
\epsfysize=35mm \epsfbox{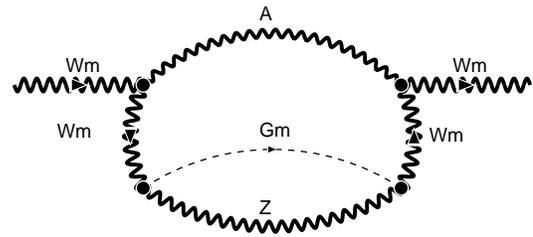}
\vspace{-.5cm}   
\caption{\label{thirdPS} Encapsulated postscript file.}
\end{figure}     
       
After all necessary files are ready, DIANA can be used to 
generate the FORM input and to execute the generated FORM program as
well.                                 
                                      
If the shapes of topologies are defined, DIANA is able to produce the
pictorial representation of diagrams, see the WEB page \\
\centerline{\footnotesize                                
\sf                                           
\mbox{http://www.physik.uni-bielefeld.de/\~{}tentukov/printing.html .}
}                                                                    
                                                                     
\noindent
Three different kinds of postscript files for the diagrams can be
produced.                                                     
                                                              
The style ``specmode.tml'' (see \cite{DIANA} p.133) contains all the
necessary function calls. Thus users of this style only need to
initialize the proper postscript driver in the environment     
\verb|initialization|.                                        
                                                              
The first driver permits the user to print all diagrams in one
file, arranging diagrams along several
rows and columns on each page, Fig. \ref{firstPS}, according to the users request.
The user must initialize the PostScript driver by means of the function 
\begin{verbatim}
   \initPostscript(  filename,
                     papersize,
                     orientation,
                     xmargin,
                     ymargin,
                     xleftmargin,
                     ncols,
                     nrows,
                     font,
                     fontsize  )
\end{verbatim}                                                 
 The parameters are:\\                                             
\verb|1.filename| -- the output file name; \\                  
\verb|2.papersize| -- one of the possible paper sizes; \\      
\verb|3.orientation| -- portrait or landscape;\\               
\verb|4.xmargin| -- both left and right margins;\\             
\verb|5.ymargin| -- both up and down margins;\\                
\verb|6.xleftmargin| -- additional left margin;\\              
\verb|7.ncols| -- number of columns per page;\\                
\verb|8.nrows| -- number of rows per page;\\                   
\verb|9.font| -- the PostScript font name;\\                   
\verb|10.fontsize| -- the PostScript font size.\\              
                             
\centerline{Example: result see Fig. \ref{firstPS}}
\begin{verbatim}              
\Begin(program,routines.rtn)  
\section(common,browser,regular)
\Begin(initialization)        
  . . .                       
 \initPostscript( pictures.ps,
                  A4,         
                  Portrait,   
                  20,         
                  20,         
                  40,         
                  2,          
                  5,          
                  Helvetica,  
                  25 )                     
  . . .                       
\End(initialization)          
  . . .                       
\End(program)                 
\end{verbatim}                
                              
The second driver prints all diagrams into one postscript file, one diagram
per page. The diagrams are printed
together with the topology and momenta flow. Such a form is
convenient not for printing, but for
investigating the diagram visually by means of some postscript
interpreter, e.g., by the ghostview, Fig. \ref{secondPS}.
To initialize this driver, the user has to define only the output file
name by means of the function \verb|\initInfoPS(filename)|.
             
\vspace*{0.5cm}            
              
\centerline{Example:}                          
\begin{verbatim}                  
\Begin(program,routines.rtn)      
\section(common,browser,regular)  
\Begin(initialization)            
  . . .          
  \initInfoPS(info.ps)            
  . . .                           
\End(initialization)              
  . . .                           
\End(program)                     
\end{verbatim}                    
                                  
The third driver can be used to create an encapsulated postscript file
containing the current diagram, Fig. \ref{thirdPS}.
To use this driver the user has to invoke the function
       \verb|\outEPS(filename,Height,font,fontsize)|. 
inside the environment \verb|output|.
If               
\verb|fontsize| = 0,
       then the particle labels will not be printed. The width of
the diagram will be defined automatically. The
       diagram will be scaled to fit the EPS bounding box 0 0 
\verb|Width| \verb|Height|.
              
\centerline{Example: result see Fig.\ref{thirdPS}}          
\begin{verbatim}                               
. . .                                          
\Begin(output,\askfilename())                  
 \outEPS( d\currentdiagramnumber().eps,        
          100,                                 
          Helvetica,                           
          15  )                                
   . . .                                       
\End(output)                                   
\end{verbatim}   
No initialization is required for this  driver.

By default, all propagators  are depicted by solid lines. To use
different kinds of lines for different particles, 
the user must define the type of the line.
At present, DIANA supports        
three types of lines: ``wavy'', suitable for vector propagators, 
``spiral'' usually used for representation of a gluon , and  ``line'' 
is just a line (full or dashed). All of them can be directed or not, and can be of
different thickness and amplitude (for ``wavy'' and ``spiral''). 
                                                                 
The syntax of the propagator description was extended as compared to the 
old one (see the DIANA 1.0 manual \\                                     
{\footnotesize
\sf                                                             
 \mbox{http://www.physik.uni-bielefeld.de/\~{}tentukov/diana\_doc.tar.gz}
}) so that it is                                                         
possible                                                                 
to define the type of the                                                
drawing line.                                                            
Let us consider, e.g., the photon propagator description:
{\small \tt                       
[A,A;a;V(num,ind:1,ind:2,vec,0);0;wavy,4,2]
}                                 
From this example we can see that before the last ``]'' 
the user can describe (optionally) 
how to draw the corresponding line. The syntax is:
\begin{verbatim}
;linetype,parameter,linewidth
\end{verbatim}
 In the above example \verb|linetype=wavy|, 
      \verb|parameter=4| and \verb|linewidth=2|

    The linetype is just an abstract type of the line. It can be one
    of the following: 
       \verb|wavy|, \verb|arrowWavy|, \verb|spiral|,   
\verb|arrowSpiral|,  \verb|line|,  \verb|arrowLine|. 
For ``wavy'' and ``spiral'' the value of the \verb|parameter|
    determines the amplitude while for ``line'' it means the type of dashing.

Another way to define a line type is to use the function 
\begin{verbatim}
  \setpropagatorline(particle, 
                     linetype, 
                     parameter, 
                     linewidth
 )
\end{verbatim}                                
in the \verb|initialization| environment, for example:
\begin{verbatim}
\Begin(initialization)
  \setpropagatorline(A,wavy,4,2)
\End(initialization)
\end{verbatim}

\end{document}